\def\mt{{m_t}}
\def\mts{{m_t^2}}

\def\shat{{\hat s}}
\def\muf{{\mu^{}_f}}
\def\mufs{{\mu^{\,2}_f}}
\def\mur{{\mu^{}_r}}

\def\msbar{{$\overline{\mbox{MS}}$}}
\def\mbar{\overline{m}}
\def\mmu{m(\mur)}
\def\mm{m(m)}

\newcommand{\GeV}{\ensuremath{\,\mathrm{GeV}}}
\newcommand{\TeV}{\ensuremath{\,\mathrm{TeV}}}

\newcommand{\lsim}{\raisebox{-0.07cm}{$\:\stackrel{<}{{\scriptstyle \sim}}\: $} }

\documentclass{PoS}

\title{
\vspace*{-2.2cm}
\begin{minipage}{\textwidth}
{\normalfont\small 
DESY 10-009, HU-EP-10/04, SFB/CPP-10-15, 
~ {\tt arXiv:1001.3987 [hep-ph]}
\hspace{\fill} January 2010}\\
\end{minipage}\\[35pt]
The top-quark's running mass}

\ShortTitle{The top-quark's running mass}

\author{\speaker{S. Moch}\\
        Deutsches Elektronen--Synchrotron, DESY   \\
        Platanenallee 6, D--15738 Zeuthen, Germany\\
        E-mail: \email{sven-olaf.moch@desy.de}
}

\author{U. Langenfeld\\
        Humboldt-Universit\"at zu Berlin, Institut f\"ur Physik \\
        Newtonstra\ss e 15, D--12489 Berlin, Germany\\
        E-mail: \email{Ulrich.Langenfeld@physik.hu-berlin.de}
}

\author{P. Uwer\\
        Humboldt-Universit\"at zu Berlin, Institut f\"ur Physik \\
        Newtonstra\ss e 15, D--12489 Berlin, Germany\\
        E-mail: \email{Peter.Uwer@physik.hu-berlin.de}
}

\abstract{
We discuss the direct determination of the running top-quark mass from 
measurements of the total cross section of hadronic top-quark pair-production.
The theory predictions in the \msbar\ scheme are 
very stable under scale variations and show rapid apparent convergence 
of the perturbative expansion. 
These features are explained by studying the underlying parton dynamics.
}

\FullConference{RADCOR 2009 - 9th International Symposium on Radiative Corrections (Applications of Quantum Field Theory to Phenomenology) ,\\
		 October 25 - 30 2009\\
		 Ascona, Switzerland}

\begin{document}

\section{Introduction}
The top-quark is the heaviest known elementary particle 
and it plays a prominent role in the physics program of Tevatron 
and the Large Hadron Collider (LHC) (see e.g.~\cite{Bernreuther:2008ju}). 
The top-quark mass is a very important parameter in fits constraining the
Standard Model (SM), i.e. giving rise to indirect limits on the mass of the
Higgs boson (see e.g.~\cite{Flacher:2008zq}).
Currently, a value of $\mt = 173.1^{+1.3}_{-1.3} \GeV$ is quoted for the mass of 
the top-quark~\cite{fnal:2009ec}. 
This amounts to an experimental uncertainty of less than 1\%. 
Due to the high mass the top-quark's width is so large that it typically decays before 
it can hadronize~\cite{Bigi:1986jk} so that mass measurements proceed via
kinematic reconstruction from the decay products and comparison to Monte Carlo simulations.
Thus, there is no immediate interpretation of the measured quantity in terms 
of a parameter of the SM Lagrangian in a specific renormalization scheme.

In order to address this issue, we have chosen the following approach. 
We start from the total cross section for hadronic top-quark pair production, 
i.e. a quantity with well-defined scheme dependence which is known to sufficient 
accuracy in perturbative Quantum Chromodynamics (QCD).
Its dependence on the top-quark mass is commonly given in the on-shell scheme, 
although it is well-known that the concept of the pole mass has intrinsic theoretical limitations 
leading, for instance, to a poorly behaved perturbative series. 
This typically implies a strong dependence of the extracted value for the
top-quark mass on the order of perturbation theory.
So-called short distance masses offer a solution to this problem. 
As we compute the total cross section as a function of the top-quark mass 
in the \msbar\ scheme~\cite{Gray:1990yh} 
we demonstrate stability of the perturbative expansion and good properties of
apparent convergence~\cite{Langenfeld:2009wd}.
In particular, this allows for the direct determination of the 
top-quark's running mass from Tevatron measurements for the total cross section~\cite{Abazov:2009ae}, 
which is of importance for global analyses of electro-weak precision data.

\section{The total cross section for top-quark-pair production}
We start by recalling the relevant formulae for the total cross section
$\sigma_{pp \to {t\bar t X}}$ of
top-quark hadro-production within perturbative QCD,
\begin{eqnarray}
  \label{eq:totalcrs}
  \sigma_{pp \to {t\bar t X}}(S,\mts) &=& 
  \sum\limits_{i,j = q,{\bar{q}},g} \,\,\,
  \int\limits_{4\mts}^{S}\,
  ds \,\, L_{ij}(s, S, \mufs)\,\,
  {\hat \sigma}_{ij}(s,\mts,\mufs)
  \, ,
\\
  \label{eq:partonlumi}
  L_{ij}(s,S,\mufs) &=& 
  {1\over S} \int\limits_s^S
  {d\shat\over \shat} 
  \phi_{i/p}\left({\shat \over S},\mufs\right) 
  \phi_{j/p}\left({s \over \shat},\mufs\right)
  \, ,
\end{eqnarray}
where $S$ denotes the hadronic center-of-mass energy squared and $\mt$ the
top-quark mass (taken to be the pole mass here). 
The standard definition for the parton luminosity $L_{ij}$ convolutes the two
parton distributions (PDFs) $\phi_{i/p}$ 
at the factorization scale $\muf$, while the partonic cross sections ${\hat \sigma}_{ij}$ 
parameterize the hard partonic scattering process. 
${\hat \sigma}_{ij}$ depends only on dimensionless ratios of $\mt$, $\muf$ and 
the partonic center-of-mass energy squared $s$.

The QCD radiative corrections for the total cross section in Eq.~(\ref{eq:totalcrs}) 
as an expansion in the strong coupling constant $\alpha_s$ 
are currently known completely at next-to-leading order 
(NLO)~\cite{Nason:1987xz} 
and, as approximation, at next-to-next-to-leading order 
(NNLO)~\cite{Moch:2008qy}. 
The latter result is based on the known threshold corrections to the partonic
cross section ${\hat \sigma}_{ij}$, 
i.e. the complete tower of Sudakov logarithms in $\beta = \sqrt{1 - 4\mt^2/s}$ and the two-loop Coulomb
corrections, i.e. powers $1/\beta^k$ (see also~\cite{Beneke:2009ye} for some recent improvements).
It also includes the complete dependence on $\muf$ and the renormalization scale $\mur$, 
both being known from a renormalization group analysis.

The parton luminosity $L_{ij}$ in Eq.~(\ref{eq:partonlumi}) is fully known to
NNLO accuracy from global fits (e.g.~\cite{Martin:2009iq,Alekhin:2009ni}).
For a fixed collider energy $S$, it is a steeply falling function of $s$. 
Thus, in the convolution Eq.~(\ref{eq:totalcrs}) 
$L_{ij}$ dominantly samples the threshold region of the underlying hard parton
scattering ${\hat \sigma}_{ij}$, which justifies the use of threshold approximations for the latter quantity.
As an upshot, the presently available perturbative corrections through NNLO lead to accurate predictions 
for the total hadronic cross section of top-quark pairs 
with a small associated theoretical uncertainty~\cite{Langenfeld:2009wd,Moch:2008qy} 
(see also e.g.~\cite{Cacciari:2008zb} for related theory improvements through threshold resummation).

\section{The top-quark mass in the \msbar\ scheme}
Colored particles in QCD are not asymptotic states of the $S$-matrix due to confinement. 
Therefore the pole mass for quarks is a poor scheme choice 
since its definition implies intrinsic uncertainties of the order of $\Lambda_{\textrm{\scriptsize QCD}}$, 
a fact that is often referred to in perturbation theory as the infrared renormalon problem.
It is well-known that short distance masses impose renormalization conditions which avoid this problem.
In a perturbative expansion in $\alpha_s$ 
the pole mass $\mt$ can be related to the running mass $\mmu$ in the \msbar\ scheme, 
\begin{equation}
  \label{eq:mpole-mbar}
  \mt = \mmu \* \left(1 + \alpha_s(\mur) d^{(1)}(\mur) + \dots \right)
  \, ,
\end{equation}
where the coefficients $d^{(l)}$ are actually known to 
three-loop order~\cite{Gray:1990yh}. 
The basic idea for the direct determination of a \msbar\ mass 
is to use the manifest dependence of the total cross section $\sigma_{pp \to {t\bar t X}}$ 
on the top-quark mass to estimate the parameter from the data for the measured cross section. 
For the pole mass $\mt$ we have 
\begin{equation}
  \label{eq:hadro-mpole}
  \sigma_{pp \to {t\bar t X}} = \alpha_s^2\, \sigma^{(0)}(\mt) + \alpha_s^3\, \sigma^{(1)}(\mt) + \dots 
  \, ,
\end{equation}
which we can convert with Eq.~(\ref{eq:mpole-mbar}) to the \msbar\ mass
$\mm$ (for simplicity abbreviated as $\mbar$) according to 
\begin{equation}
  \label{eq:hadro-mbar}
  \sigma_{pp \to {t\bar t X}} = 
    \alpha_s^2\, \sigma^{(0)}(\mbar) 
    + 
    \alpha_s^3\, \left( 
      \sigma^{(1)}(\mbar) 
      + \mbar\, d^{(1)} \partial_m \sigma^{(0)}(m) \biggr|_{m=\mbar}
      \right) 
    + 
    \dots
  \, ,
\end{equation}
where the coefficients $d^{(l)}$ 
have to be evaluated for $\mu_r = \mbar$ (corresponding to the scale of $\alpha_s$).
In Eqs.~(\ref{eq:mpole-mbar})--(\ref{eq:hadro-mbar}) we have confined ourselves here for brevity to NLO 
(see~\cite{Langenfeld:2009wd} for the formalism through NNLO).

\begin{figure}[t!]
\centering
\vspace*{10mm}
    {
        \includegraphics[bb = 30 55 540 500, scale = 0.345]{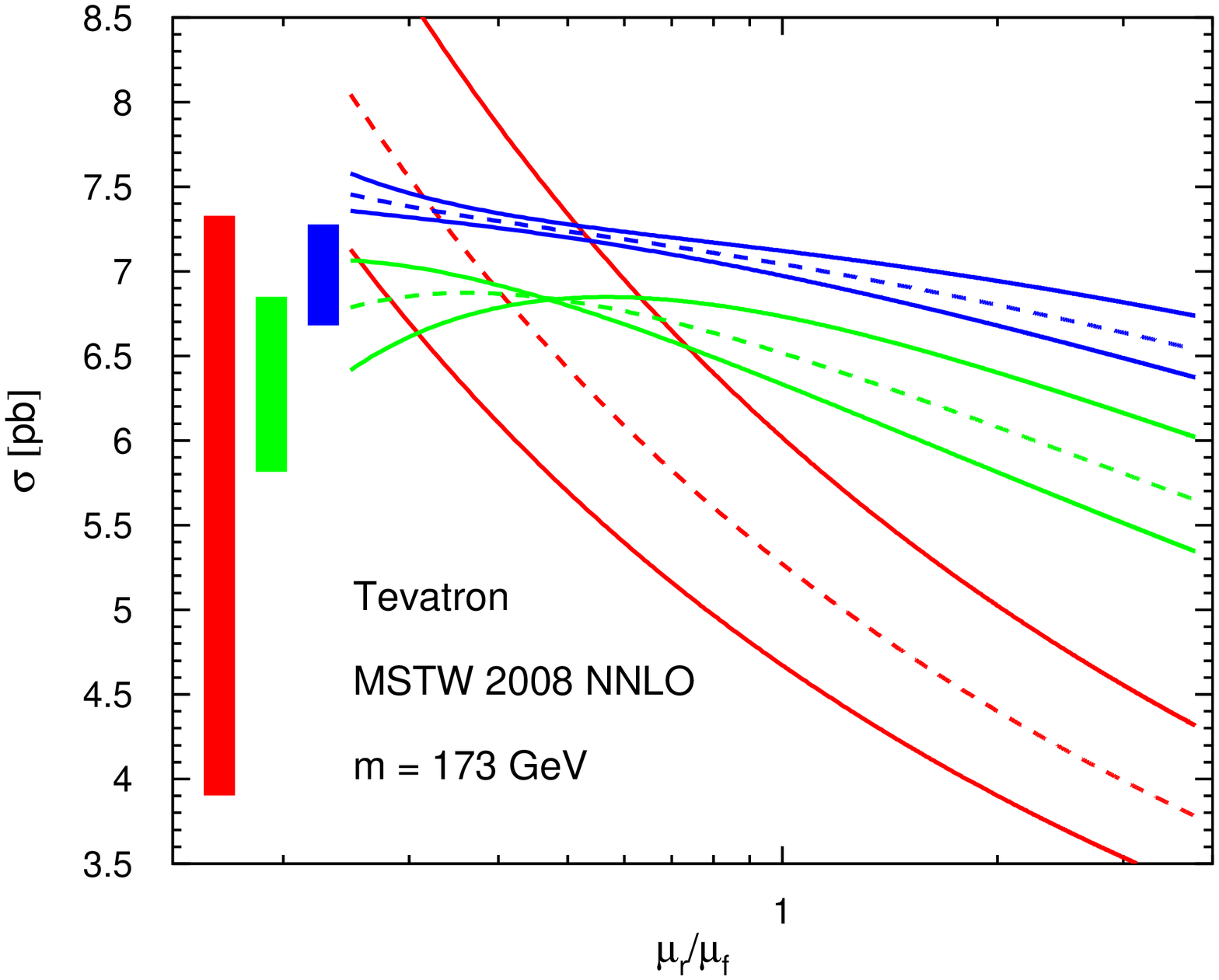}
    }
\hspace*{10mm}
    {
        \includegraphics[bb = 30 55 540 500, scale = 0.345]{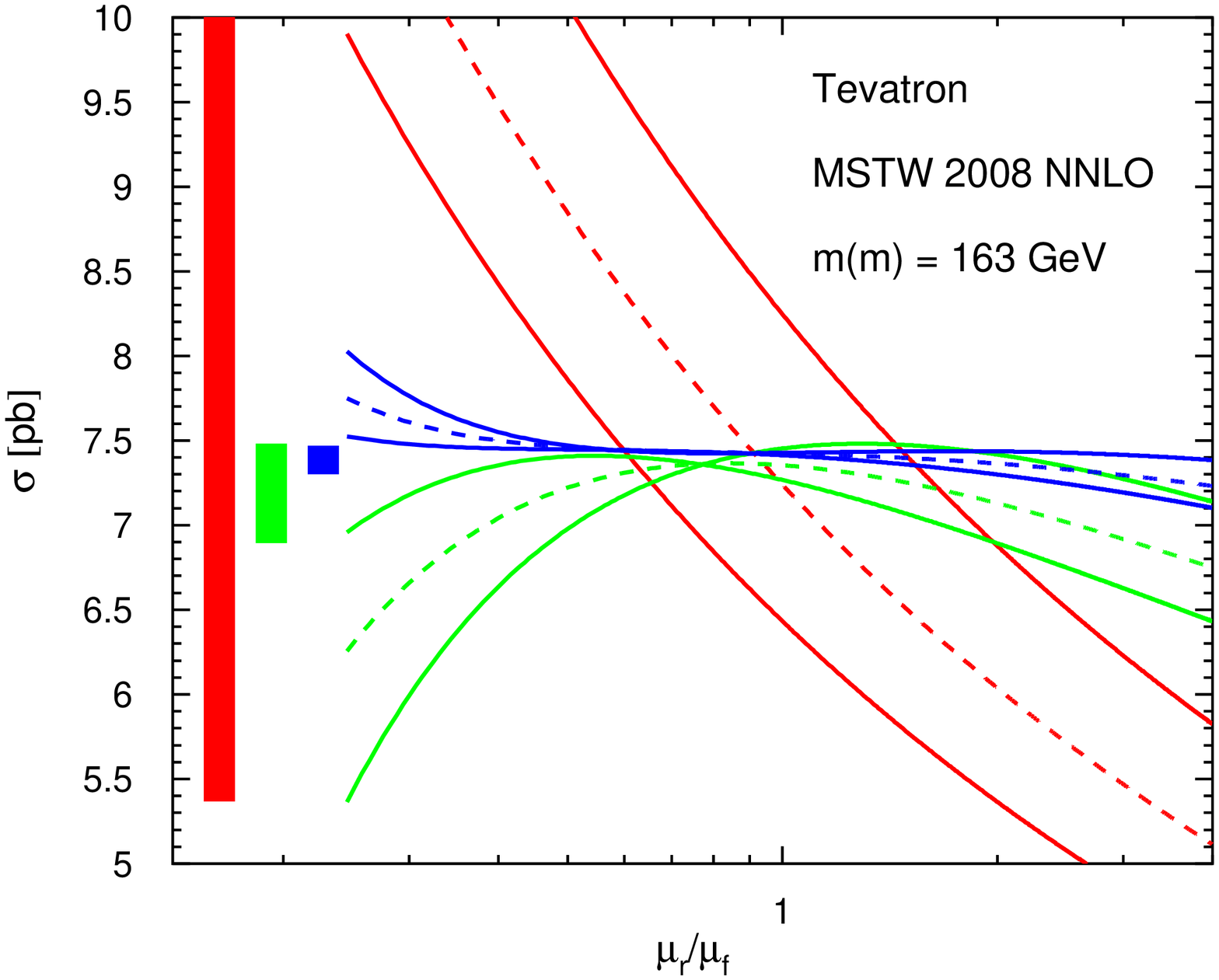}
    }
\hspace*{3mm}
\caption{\small
  \label{fig:msbarscale}
 The scale dependence of the total cross section at Tevatron 
 with $\sqrt{S} = 1.96\TeV$ with MSTW 2008 PDF set~\cite{Martin:2009iq}.
 The top-quark mass is taken 
 in the on-shell scheme at $\mt = 173\GeV$ (left) and
 in the \msbar\ scheme at $\mbar = 163\GeV$ (right) at LO (red), NLO (green) and
 approximate NNLO (blue).
 The dashed lines denote the choice $\muf = \mt$ (left) and $\muf = \mbar$ (right) for the factorization scale, 
 the solid lines the maximal deviations for 
 $\mur \in [\mt/2, 2\mt]$ and $\muf = \mt/2, \mt$ and $2\mt$ (left) and 
 $\mur \in [\mbar/2, 2\mbar]$ and $\muf = \mbar/2, \mbar$ and $2\mbar$ (right).
 The vertical bars indicate the size of the scale variation in the standard
 range $[\mt/2, 2\mt]$ (left) and $[\mbar/2, 2\mbar]$ (right).
}
\end{figure}

Eq.~(\ref{eq:hadro-mbar}) gives a direct handle on the running mass at large scales.
To illustrate the phenomenological implications for predictions at hadron colliders, 
we plot in Fig.~\ref{fig:msbarscale} the scale dependence of the total cross section 
at the various orders in perturbation theory.
For Tevatron with $\sqrt{S} = 1.96\TeV$ (and using the MSTW 2008 PDF set~\cite{Martin:2009iq}), 
we compare the on-shell scheme with a pole mass of $\mt = 173 \GeV$ with the 
corresponding predictions for a running mass with a value of $\mbar = 163 \GeV$.
For the computation of the total cross section in the on-shell scheme, 
we choose three (fixed) values for the factorization scale $\muf = \mt/2, \mt$ and $2 \mt$ 
and, likewise $\muf = \mbar/2, \mbar$ and $2 \mbar$ for the \msbar\ scheme.
The vertical bands on the left in Fig.~\ref{fig:msbarscale} denote the maximum and the minimum values for a
variation of $\mur \in [\mt/2, 2\mt]$ (and, respectively, $\mur \in [\mbar/2, 2\mbar]$) for the three choices of $\muf$.

In general, we observe in both schemes a reduced scale dependence as we increase the order 
of perturbation theory, i.e. a reduced theoretical uncertainty. 
Also, we do observe apparent convergence of the expansion upon including successive orders in $\alpha_s$. 
For the on-shell scheme, however, the higher order corrections are quite sizable, 
${\cal O}(30 \%)$ at NLO and another ${\cal O}(10 \%)$ at NNLO at the central value $\mur = \muf = \mt$.
For the running \msbar\ mass on the other hand 
both NLO and NNLO corrections are negligible for the choice $\mur = \muf = \mbar$.
Remarkably, in the \msbar\ scheme we do find even greater stability with respect to
scale variations, which at NLO and NNLO is reduced by more than 
a factor of two compared to the results in the pole mass scheme.
Similar results and conclusions have been found for top-quark pair
production at LHC, see~\cite{Langenfeld:2009wd}, 
although the improvement is slightly less distinct than at Tevatron.

In order to address the underlying parton dynamics of relevance for the two mass schemes 
it is instructive to consider the total parton cross sections ${\hat \sigma}_{ij}$, 
i.e. the equivalent expression of Eq.~(\ref{eq:hadro-mbar}) for the individual partonic channels. 
As a matter of fact, it turns out, that a result completely analogous to Eq.~(\ref{eq:hadro-mbar}) can be derived.
To NLO this is true because the boundary term in the conversion $\mt \to \mbar$ 
from the convolution integral in Eq.~(\ref{eq:totalcrs}) vanishes, 
so that we can apply Eq.~(\ref{eq:hadro-mbar}) with the simple replacement $\sigma \to {\hat \sigma}_{ij}$.

\begin{figure}[t!]
\centering
\vspace*{10mm}
    {
        \includegraphics[bb = 55 95 535 735, scale = 0.32, angle = 270]{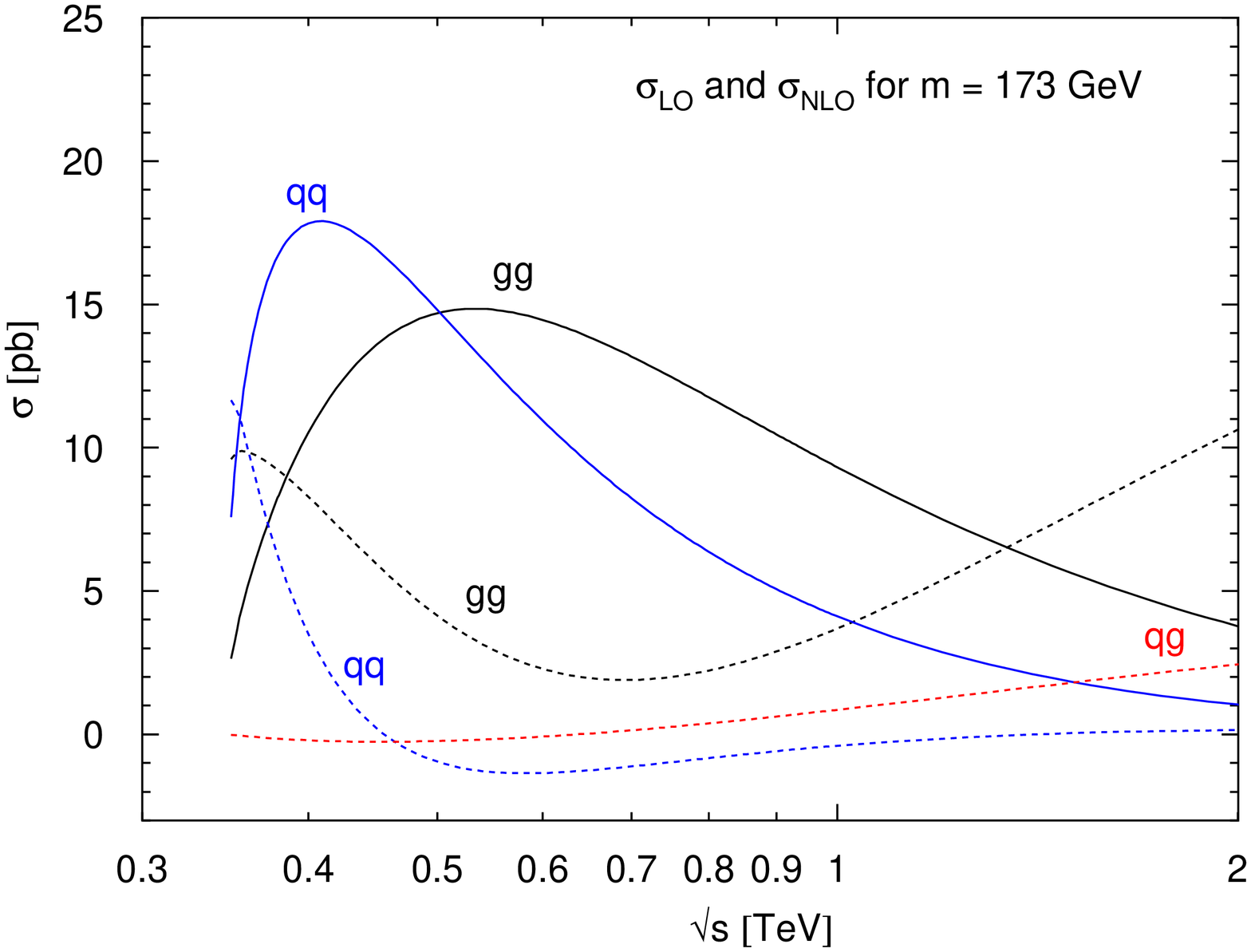}
    }
\hspace*{1mm}
    {
        \includegraphics[bb = 55 95 535 735, scale = 0.32, angle = 270]{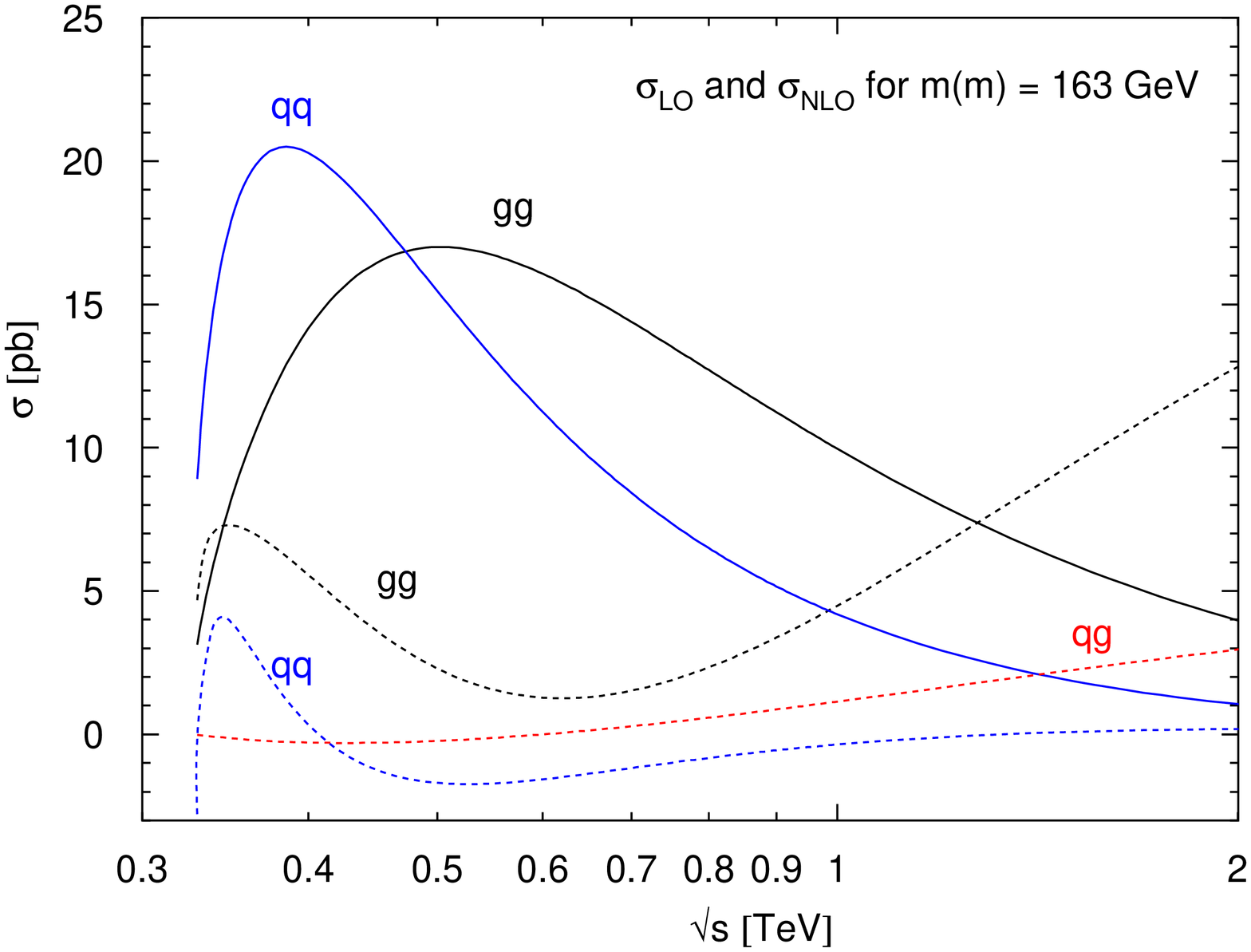}
    }
\hspace*{3mm}
\caption{\small
  \label{fig:msbar-partonic}
 The parton cross section for the channels $q\bar q$, $gg$ and $qg$ 
 at the scale $\muf = \mur = m$ in the on-shell scheme 
 for $\mt = 173\GeV$ (left) and 
 in the \msbar\ scheme for $\mbar = 163\GeV$.
 Solid lines denote the ${\cal O}(\alpha_s^2)$ (LO)
 and dashed lines the ${\cal O}(\alpha_s^3)$ contributions (NLO).
 The energy range corresponds to Tevatron with $\sqrt{S} = 1.96\TeV$ 
 and the value of $\alpha_s$ to MSTW 2008 PDF set~\cite{Martin:2009iq}.
}
\end{figure}

In Fig.~\ref{fig:msbar-partonic} we plot ${\hat \sigma}_{ij}$ 
in both schemes, i.e. the on-shell scheme with $\mt = 173 \GeV$ and the 
\msbar\ scheme with a running mass $\mbar = 163 \GeV$ as a function of the
partonic center-of-mass energy $s$. 
The energy range is selected to match the discussion for the Tevatron around Fig.~\ref{fig:msbarscale}.
Of course, the Born cross sections remain largely unchanged the only
difference in the \msbar\ case being the slightly smaller numerical value of the
mass (hence, larger cross sections).
At NLO, the perturbative corrections in the on-shell scheme for the channels
$q\bar q$ and $gg$ clearly display the well-known large logarithmic corrections near threshold.  
This is not the case for the \msbar\ scheme, which 
exhibits a much reduced sensitivity to the threshold region.
Due to the terms $\sim \partial_m {\hat \sigma}_{ij}^{(0)}$ in the partonic equivalent of Eq.~(\ref{eq:hadro-mbar}), 
the NLO corrections are sizably reduced and the Sudakov logarithms are
numerically compensated to a large extent.
The $qg$-channel is new at NLO, thus it does not receive any modification 
under scheme transformations at this order.

The parton cross sections of Fig.~\ref{fig:msbar-partonic} enter 
the convolution with the parton luminosity $L_{ij}$ as given in Eq.~(\ref{eq:totalcrs}).
To that end, recall that the hadronic cross section at Tevatron almost saturates already 
for partonic center-of-mass energies $\sqrt{s} \lsim 600 \GeV$.
A detailed treatment of the threshold region e.g. in
Fig.~\ref{fig:msbar-partonic} also needs to incorporate $t\bar t$ bound state effects 
which requires the application of non-relativistic QCD including an all-order resummation of 
Coulomb corrections, see~\cite{Hagiwara:2008df}. 

As an upshot, the parton level studies of the \msbar\ case in Fig.~\ref{fig:msbar-partonic}
provide us with a detailed understanding of the excellent apparent convergence 
and scale stability seen in Fig.~\ref{fig:msbarscale}.
In a direct comparison to data~\cite{Abazov:2009ae}, this leads to very stable results 
for the extracted mass parameter. 
At LO, NLO, and NNLO values of 
$\mbar = 159.2^{+3.5}_{-3.4} \GeV$,
$\mbar = 159.8^{+3.3}_{-3.3} \GeV$ and 
$\mbar = 160.0^{+3.3}_{-3.2} \GeV$ are determined in~\cite{Langenfeld:2009wd},
where the errors reflect the quoted experimental uncertainty for the total
cross section.
In contrast, the on-shell scheme predictions would return rather different
results at the higher orders.
Converting the best estimate for the running mass (i.e. the NNLO value) back to the on-shell mass by inverting Eq.~(\ref{eq:mpole-mbar})
leads to a pole mass value of $\mt = 168.9^{+3.5}_{-3.4} \GeV$.
Within errors, the result is consistent with the direct measurements, 
although as mentioned above, concerns have been raised to interpret the 
quoted value~\cite{fnal:2009ec} of $\mt = 173.1^{+1.3}_{-1.3} \GeV$ as a pole mass. 
Since the experimental analysis is based to large extend on leading-order Monte Carlo prescriptions, 
additional efforts are needed to study the detailed scheme dependence, 
see e.g.~\cite{Hoang:2008yj} 
for the renormalization group flow for heavy quark masses.

\section{Summary}

We have computed the total cross section for top-quark pair production with
the \msbar\ mass definition for the top-quark~\cite{Langenfeld:2009wd}.
The approximate NNLO predictions exhibit a greatly improved pattern of apparent convergence for the
perturbative expansion and very good stability with respect to scale variations.
Comparison with experimental data has lead to a best estimate for the running 
mass of $\mbar=160.0^{+3.3}_{-3.2} \GeV$, 
which is the first direct determination of $\mm$ from top-quark pair-production.
The corresponding value for the pole mass of $\mt = 168.9^{+3.5}_{-3.4} \GeV$
is consistent with current world average~\cite{fnal:2009ec}, $\mt = 173.1^{+1.3}_{-1.3} \GeV$.

Altogether, our approach~\cite{Langenfeld:2009wd,Moch:2008qy} 
provides reliable approximate NNLO predictions for the total cross section for
top-quark pair production and stable values for the top-quark's running 
mass.


\begin{thebibliography}{10}

\bibitem{Bernreuther:2008ju}
W. Bernreuther,
\newblock J. Phys. G35 (2008) 083001, arXiv:0805.1333
\newblock 
\\
J.R. Incandela et~al.,
\newblock Prog. Part. Nucl. Phys. 63 (2009) 239, arXiv:0904.2499
\newblock 

\bibitem{Flacher:2008zq}
H. Flacher et~al.,
\newblock Eur. Phys. J. C60 (2009) 543, arXiv:0811.0009
\newblock 

\bibitem{fnal:2009ec}
Tevatron Electroweak Working Group,
\newblock (2009), arXiv:0903.2503
\newblock 

\bibitem{Bigi:1986jk}
I.I.Y. Bigi et~al.,
\newblock Phys. Lett. B181 (1986) 157
\newblock 

\bibitem{Gray:1990yh}
N. Gray et~al.,
\newblock Z. Phys. C48 (1990) 673
\newblock 
\\
K.G. Chetyrkin and M. Steinhauser,
\newblock Nucl. Phys. B573 (2000) 617, hep-ph/9911434
\newblock 
\\
K. Melnikov and T.v. Ritbergen,
\newblock Phys. Lett. B482 (2000) 99, hep-ph/9912391
\newblock 

\bibitem{Langenfeld:2009wd}
U. Langenfeld, S. Moch and P. Uwer,
\newblock Phys. Rev. D80 (2009) 054009, arXiv:0906.5273
\newblock 

\bibitem{Abazov:2009ae}
D0, V.M. Abazov et~al.,
\newblock (2009), arXiv:0903.5525
\newblock 

\bibitem{Nason:1987xz}
P. Nason, S. Dawson and R.K. Ellis,
\newblock Nucl. Phys. B303 (1988) 607
\newblock 
\\
W. Beenakker et~al.,
\newblock Phys. Rev. D40 (1989) 54
\newblock 
\\
W. Bernreuther et~al.,
\newblock Nucl. Phys. B690 (2004) 81, hep-ph/0403035
\newblock 
\\
M. Czakon and A. Mitov,
\newblock Nucl. Phys. B824 (2010) 111, arXiv:0811.4119
\newblock 

\bibitem{Moch:2008qy}
S. Moch and P. Uwer,
\newblock Phys. Rev. D78 (2008) 034003, arXiv:0804.1476
\newblock 
\\
S. Moch and P. Uwer,
\newblock Nucl. Phys. Proc. Suppl. 183 (2008) 75, arXiv:0807.2794
\newblock 

\bibitem{Beneke:2009ye}
M. Beneke et~al.,
\newblock (2009), arXiv:0911.5166
\newblock 

\bibitem{Martin:2009iq}
A.D. Martin et~al.,
\newblock Eur. Phys. J. C63 (2009) 189, arXiv:0901.0002
\newblock 

\bibitem{Alekhin:2009ni}
S. Alekhin et~al.,
\newblock (2009), arXiv:0908.2766
\newblock 

\bibitem{Cacciari:2008zb}
M. Cacciari et~al.,
\newblock JHEP 09 (2008) 127, arXiv:0804.2800
\newblock 

\bibitem{Hagiwara:2008df}
K. Hagiwara, Y. Sumino and H. Yokoya,
\newblock Phys. Lett. B666 (2008) 71, arXiv:0804.1014
\newblock 
\\
Y. Kiyo et~al.,
\newblock Eur. Phys. J. C60 (2009) 375, arXiv:0812.0919
\newblock 

\bibitem{Hoang:2008yj}
A.H. Hoang et~al.,
\newblock Phys. Rev. Lett. 101 (2008) 151602, arXiv:0803.4214
\newblock 
\\
A.H. Hoang and I.W. Stewart,
\newblock Nucl. Phys. Proc. Suppl. 185 (2008) 220, arXiv:0808.0222
\newblock 

\end{thebibliography}

\end{document}